\documentclass[aps,prl,showpacs,floatfix,preprintnumbers,nofootinbib,
               amssymb,amsmath,twocolumn]{revtex4}
\usepackage{subfigure}
\usepackage{graphicx,bm,color}
\usepackage{amssymb}

\begin{document}

\title{Net Charge Fluctuations as a signal of QGP from\\ 
       Polyakov$-$Nambu$-$Jona-Lasinio model}

\author{Abhijit Bhattacharyya}
\email{abphy@caluniv.ac.in}
\affiliation{Department of Physics, University of Calcutta,
             92, A.P.C Road, Kolkata-700009, INDIA}
\author{Supriya Das}
\email{supriya@bosemain.boseinst.ac.in}
\author{Sanjay K. Ghosh}
\email{sanjay@bosemain.boseinst.ac.in}
\author{Sibaji Raha}
\email{sibaji@bosemain.boseinst.ac.in}
\author{Rajarshi Ray}
\email{rajarshi@bosemain.boseinst.ac.in}
\author{Kinkar Saha}
\email{saha.k.09@gmail.com}
\author{Sudipa Upadhaya}
\email{sudipa.09@gmail.com}
\affiliation{Center for Astroparticle Physics \&
Space Science, Block-EN, Sector-V, Salt Lake, Kolkata-700091, INDIA 
 \\ \& \\ 
Department of Physics, Bose Institute, \\
93/1, A. P. C Road, Kolkata - 700009, INDIA}

\begin{abstract}
We report the first model study of  the net charge fluctuations in terms
of $D-measure$ within the framework of the
Polyakov$-$Nambu$-$Jona-Lasinio model. Net charge fluctuation is
estimated from the charge susceptibility evaluated using PNJL model. A
parameterization of the freeze-out curve has been used to obtain $D$ as
a function of $\sqrt s$. We have discussed our results vis-a-vis recent
experimental findings from ALICE collaboration.
\end{abstract}
\maketitle

Strongly interacting matter at very high temperatures and densities is
expected to undergo a transition from confined state of colored charges,
the hadronic phase with broken chiral symmetry to a partonic phase
in which chiral symmetry is restored and/or quarks are deconfined
\cite{Hildegard}. A good understanding of this transition is relevant
for studies in the fundamental interactions in particle physics, as well
as for the physics of early universe and neutron stars. Thus it has
become an issue of great interest in recent years, both theoretically
and experimentally \cite{review}. To this end, it is essential to
identify unambiguous signals which would establish the formation of a
quark-gluon plasma (QGP). One such viable signal is the fluctuations of
net electric charge $Q$ \cite{JK,KJ}. It has been argued that this
fluctuation is proportional to the square of the electric charge which
takes up distinct values for the hadronic and QGP phases. While the unit
of $Q$ in the hadronic phase is 1, that in the QGP phase is 1/3. This
may result in the fluctuation in net charge to vary with the change of
phase, with the net charge remaining unaffected. Such fluctuations in
heavy ion collision experiments are measurable via event-by-event (EbE)
analysis \cite{Jeon,Stock,Mrowczynski,Stephanov,Doring}, where one
single event corresponds to a set of innumerable particles produced in
a single collision of relativistic nuclei. This method deals with
measurement of a given observable on an EbE basis and study of
fluctuations over an ensemble of events.

{\it Measuring charge fluctuations:} To reduce systematic uncertainties
of measurable quantities in heavy-ion experiments, it is useful to
consider suitable ratios of quantities that are expected to have similar
systematic behavior. Here for measuring charge fluctuations a suitable
observable could be the ratio

\begin{equation}
 F~=~\frac{Q}{N_{ch}},
\end{equation}

\noindent
of net charge $Q$ to total charge $N_{ch}$. One could also consider a
ratio $R$ defined as,

\begin{equation}
R~=~\frac{N_+}{N_-}~=~\frac{1+F}{1-F}\\
\end{equation}

\noindent
Here, if one uses the approximation
$\left<N_{ch}\right> >> \left<Q\right>$, then
$\left<\delta R^2\right> \simeq 4\left<\delta F^2\right>$ where,

\begin{equation}
 \left<\delta F^2\right>~
      \simeq~\frac{\left<\delta Q^2\right>}{\left<N_{ch}\right>^2}
\end{equation}

\noindent
So the signal in fluctuations of $F$ is amplified four times in the
fluctuations of $R$. Now $R$ is related to ratio of net charge
fluctuation to $\left<N_{ch}\right>$ via a quantity $D$, defined as:

\begin{equation}
 D~=~\left<N_{ch}\right>\left<\delta R^2\right>
  ~=~4\frac{\left<\delta Q^2\right>}{\left<N_{ch}\right>}
  ~=~4\frac{\chi_Q}{n_{ch}/T^3}
\label{eqD}
\end{equation}

\noindent
where $n_{ch}=\left<N_{ch}\right>/V$ is the total charge density.
$n_{ch}$ may be obtained by adding the contribution from particle and
anti-particle distributions while the dimensionless charge
susceptibility $\chi_Q$ may be obtained from the pressure $P$ of the
system as,

\begin{equation}
 \chi_{Q}\left(T,\frac{\mu_{Q}}{T}\right)
  ~=~\frac{\partial^2}{\partial(\frac{\mu_{Q}}{T})^2}
     \left(\frac{P\left(T,\frac{\mu_{Q}}{T}\right)}
     {T^4}\right),
 \label{eqtaylor}
\end{equation}

\noindent
where $\mu_Q$ is the electric charge chemical potential.

This definition is useful over that of the ratio fluctuations as the
general form of the latter may be quite complicated. As mentioned
above that only if $\left<N_{ch}\right>~>>~\left<Q\right>$ the
simplified relation exists as given above.  In general this condition
is expected to be satisfied only for a large $T$ and very small
$\mu_B$. Once the $\mu_B$ becomes large the net charge $Q$ increases
and the approximation would fail. However the definition of $D$ is
quite general and holds even if not so simply related to the ratio
fluctuation.

 A simple
estimate of $D$ was made considering the hadronic phase to be composed
of pion gas and the partonic phase as computed in the Lattice Gauge
Theory \cite{JK}. This gave the value of $D$ to be $\sim 4$ for the
hadronic phase and  $\sim 1$ for QGP phase. 

The first measurement of $D$ in experiments have been reported
recently by the ALICE collaboration \cite{alice}. They have obtained
the net charge fluctuations in a rapidity window
$0.2 < \Delta \eta < 1.6$ using center of mass energy
$\sqrt{s} =$ 2.76 TeV in Pb-Pb run for different centralities. The
$D$ for increasing $\Delta \eta$ continues to fall and is just short of
the saturation region as expected from UrQMD simulation results
\cite{Jeon}. The analysis of $D$ and its variation with center of mass
energies in the range of $19.6 {\rm GeV} < \sqrt{s} < 200 {\rm GeV}$
has also been presented in the same report. The data used were obtained
by the STAR collaboration \cite{star}.

{\it Results in PNJL model:} Here we report on the study of net charge
fluctuations
in terms of the $D-measure$ using the Polyakov loop enhanced
Nambu$-$Jona-Lasinio (PNJL) model. In this model the quarks and Polyakov
loop fields are the basic degrees of freedom. The quarks while
interacting with the Polyakov loop also has a four-fermi
self-interaction. Similarly the Polyakov loop fields interact via a
Landau-type potential. The details of the model used for a 2 flavor
system may be found in \cite{ratti1,ray3}. The extension to 2+1 flavor
system has been done in \cite{deb1}. Detailed studies of fluctuations
and correlations of various conserved charges were performed with the
PNJL model both for 2 flavor \cite{ray1,ray2} and for 2+1 flavor
\cite{deb2,lahiri,sarbani} systems.

\begin{figure}[!htb]
{\includegraphics[scale=0.25,angle=270]{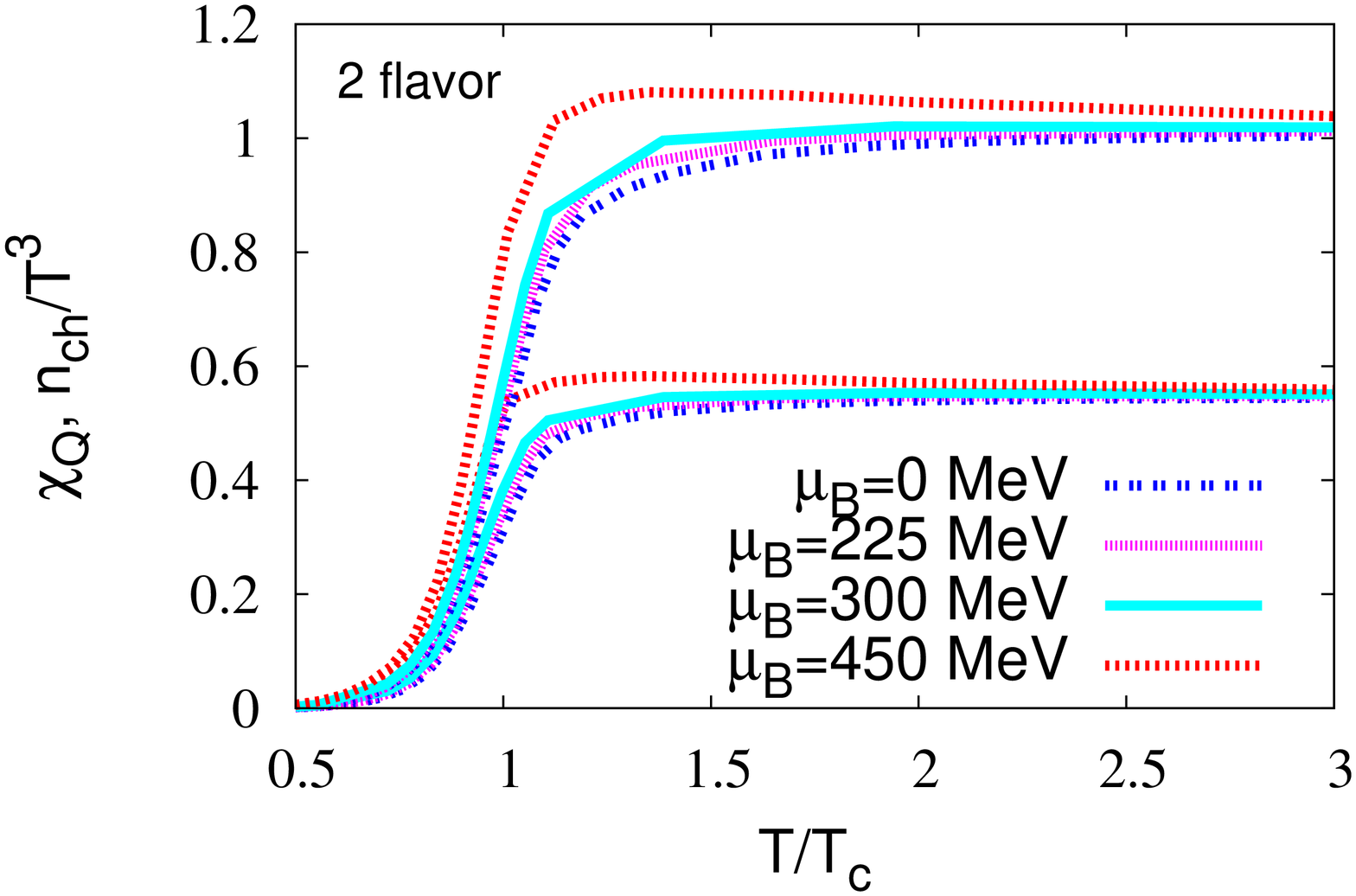}
 \includegraphics[scale=0.25,angle=270]{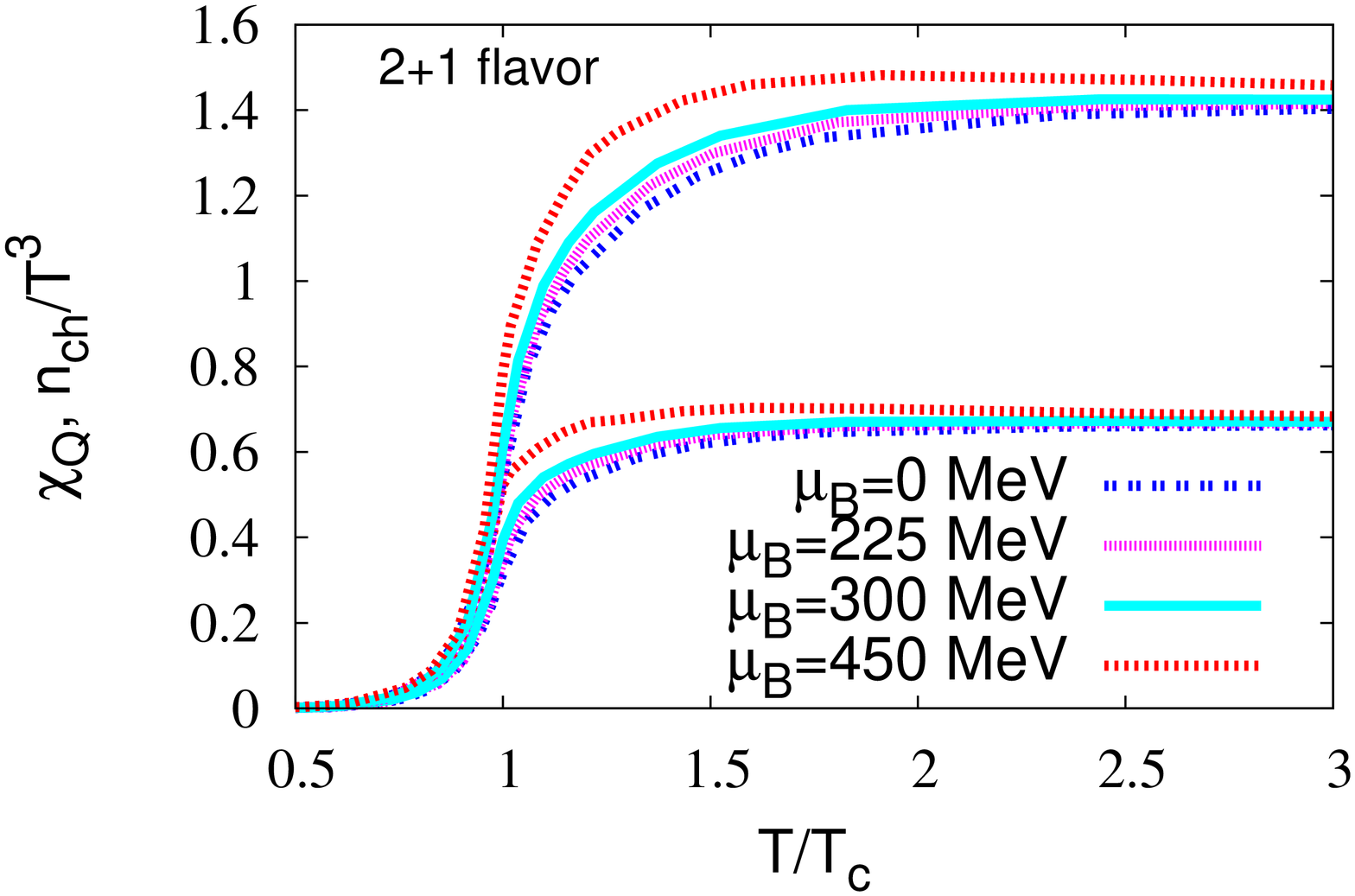}}
 \caption{(color online)Variation of $\chi_Q$ and $n_{ch}/T^3$
	 (lower and upper set of curves) with $T/T_c$ for
different values of $\mu_B$ for 2 flavor (upper panel) and 2+1 flavor
(lower panel).} 
\label{fg.susc}
\end{figure}

To compute $D$ we evaluate $\chi_Q$ and $n_{ch}$ using the PNJL model.
The method of obtaining $\chi_Q$ is quite standard as has been
discussed by us earlier \cite{ray1}. On the other hand $n_{ch}$ are to
be calculated from the quark distribution functions as they appear in
the PNJL model.

The behavior of $\chi_Q$ and $n_{ch}/T^3$ with $T/T_c$ are shown in
Fig.\ref{fg.susc} for various values of the baryon chemical potential
$\mu_B$ and for the cases of 2 flavor and 2+1 flavor systems. Here
$T_c$ is the crossover temperature at the corresponding values of
$\mu_B$. The quantities under consideration show qualitatively similar
behavior. There is a sharp rise close to $T/T_c\sim1$ from a negligibly
small value for low temperatures. This is followed by a saturation at
high temperatures close to the corresponding values for massless free
quarks. The values of $\chi_Q$ and $n_{ch}/T^3$, at any $T/T_c$ are
larger for larger $\mu_B$. However as $T$ increases they all tend
towards the limit of free gas at $\mu_B=0$.

\begin{figure}[!htb]
 {\includegraphics[scale=0.53]{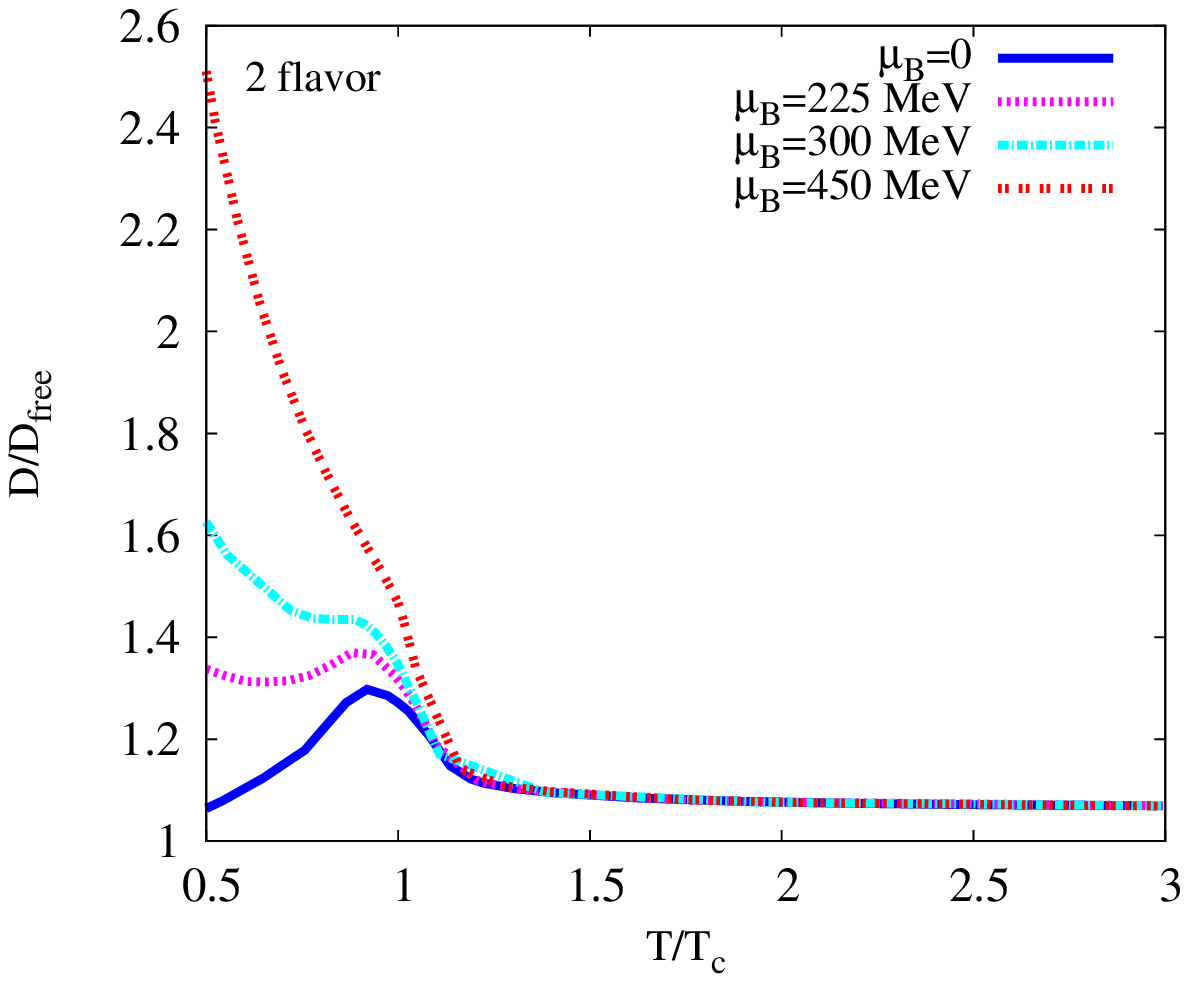}
 \includegraphics[scale=0.53]{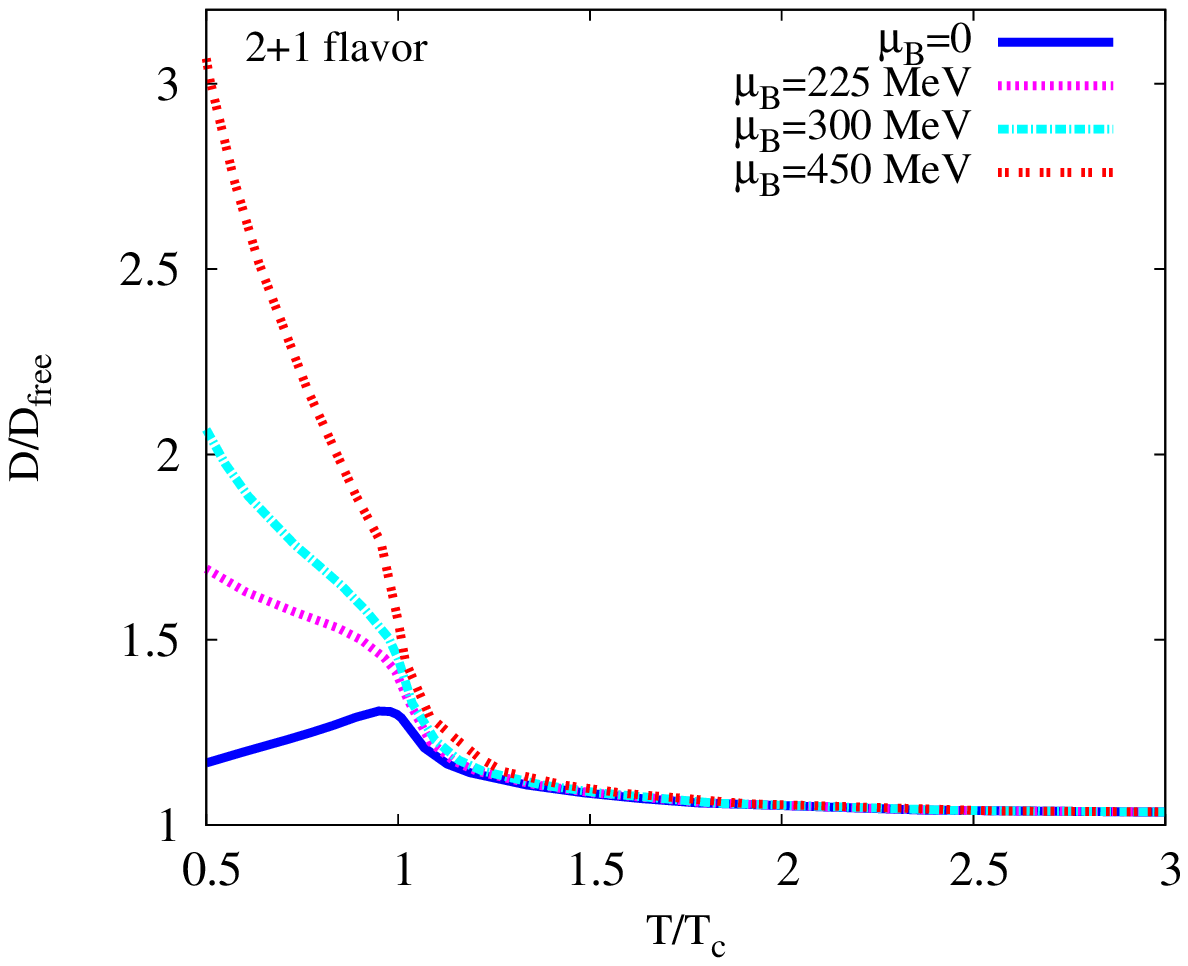}}
 \caption{(color online)Variation of D with 
$T/T_c$ for different values of $\mu_B$ around $\mu_Q=0$}. 
\label{fg.dtemp}
\end{figure}

We now consider the quantity $D/D_{free}$ and study its
temperature and density variations. Here $D_{free}(T,\mu_B)$ is
the temperature and chemical potential dependent limit of $D(T,\mu_B)$
for a free massless gas of quarks. In Fig.\ref{fg.dtemp}, we show the
variation of $D/D_{free}$ as a function of $T/T_c$ for both $2$ and
$2+1$-flavor cases. We have chosen four representative values of
$\mu_B$. It is observed that $D$ always remains above its free field
limit and approaches this limit at high enough $T$. The sharpest
transition occurs near $T=T_c$. Now both $\chi_Q$ and $n_{ch}$ are
smaller than their corresponding free field limit below $T=T_c$ for
any $\mu_B$. Therefore $D > D_{free}$ implies that $n_{ch}$ is much
more suppressed than $\chi_Q$ in the confined phase. This effect seems
to be much more prominent as $\mu_B$ is increased. In passing, let us
mention that $\left<Q\right> / \left<N_{ch}\right>$ in our
studies varied from a value of 0.025 for high $T$ and low $\mu_B$ to
0.3 at the other extreme. 

\begin{figure}[!htb]
 {\includegraphics[scale=0.25,angle=270]{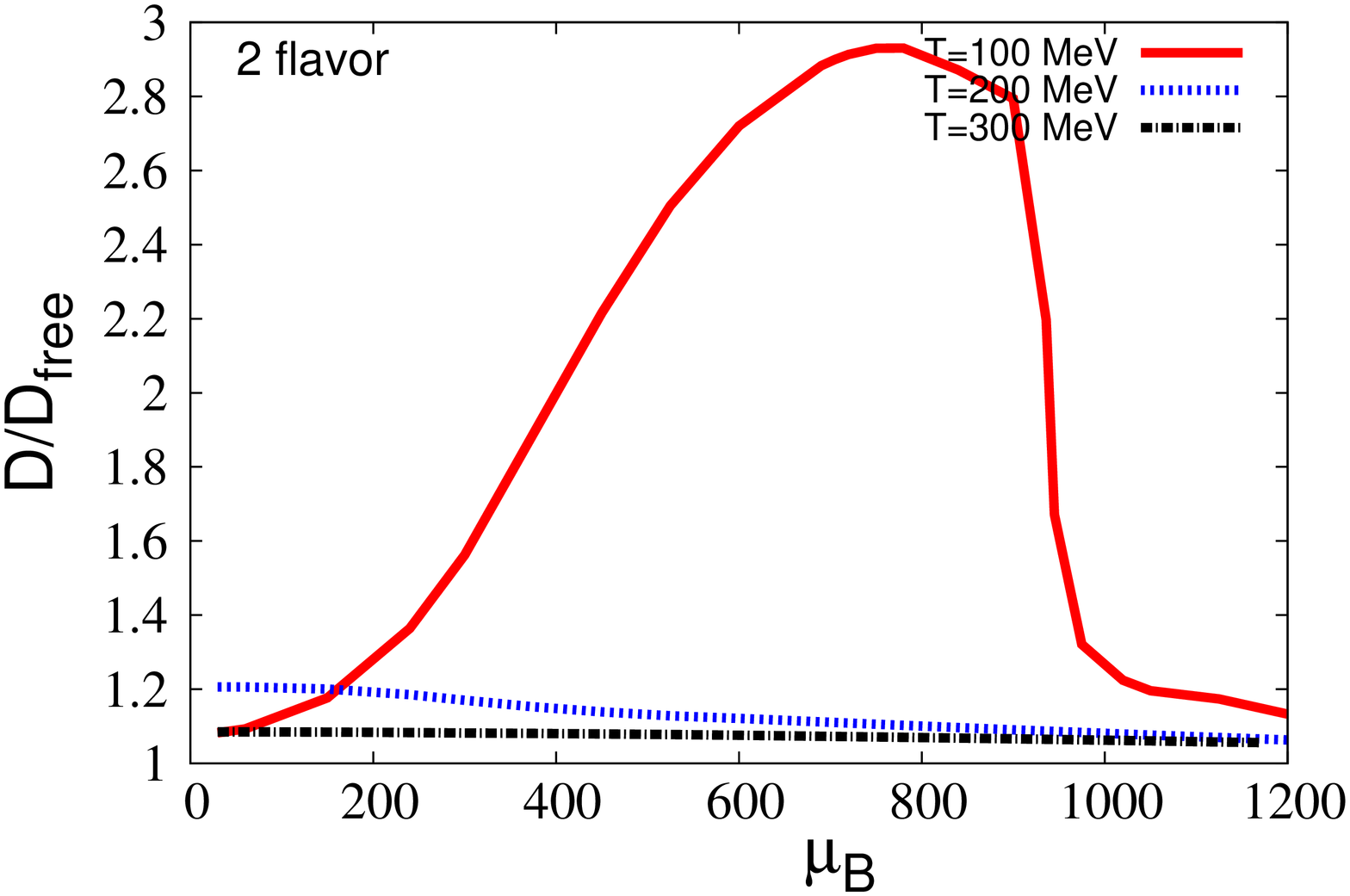}
 \includegraphics[scale=0.25,angle=270]{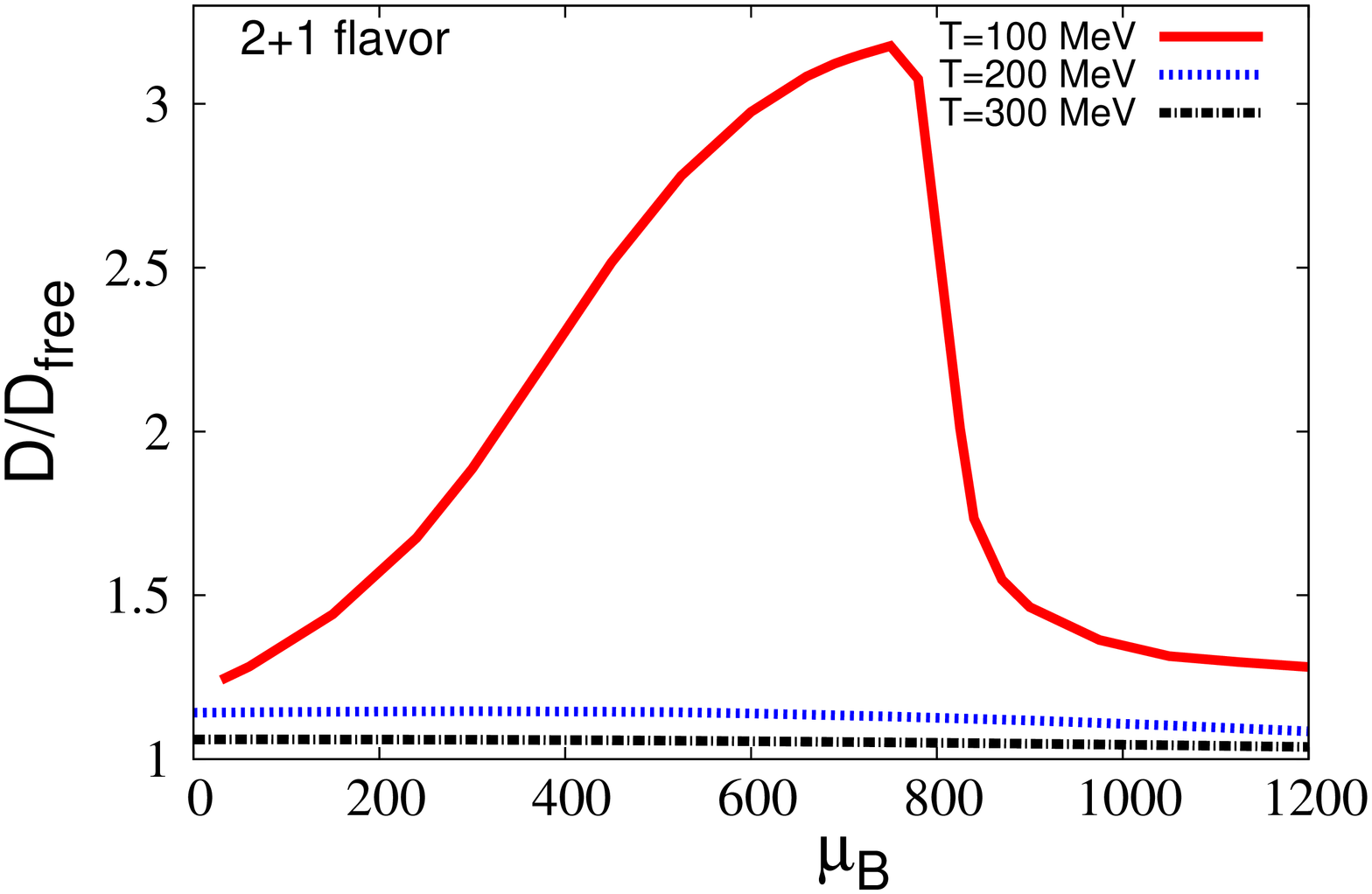}}
\caption{(color online) Variation of $D/D_{free}$ with $\mu_B$ for
three values of T around $\mu_B=0$ for 2 flavors (upper panel) and
2+1 flavors (lower panel)}. 
\label{fg.dmuB2flv}
\end{figure}

{\it Connection with heavy-ion collision experiments:} It would thus be
interesting to check what happens with further increase
of $\mu_B$. The variation of $D/D_{free}$ with $\mu_B$ at different
temperatures are shown in Fig. \ref{fg.dmuB2flv}. Again we find $D$
to remain above its free field limit for all $T$ and $\mu_B$. For a
lower temperature $\sim 100$ MeV, there is an initial rise and a
subsequent fall with increasing $\mu_B$. This non-monotonic behavior at
low temperatures may pose a problem in using $D$ as a direct indicator
of the phase of strongly interacting matter.

Nonetheless, with an input of temperature and chemical potential from
particle multiplicities at the freeze-out surface in heavy ion collision
experiments, one may study the expected nature of $D$ for different
experimental conditions. The independent thermodynamic variables in
the PNJL model are $T$, $\mu_B$, $\mu_Q$, and $\mu_S$, where the latter
is the strangeness chemical potential. Parameterization of the freeze-out
conditions as a function of $\sqrt s$ are available in the literature
(see e.g. \cite{Redlich, Andronic}). For a given set of the
thermodynamic variables we found the variations in $\sqrt s$ are
within 10$\%$ for different parameterizations. Here we choose the
parameterization in \cite{Redlich} to model the freeze-out curve as:

\begin{eqnarray}
 T(\mu_B)~=~a-b\mu_B^2-c\mu_B^4 \\ 
 \mu_{B,Q,S}(\sqrt s)~=~\frac{d}{1+e\sqrt s} 
\label{eq.energy}
\end{eqnarray}

\noindent
where, $a=0.166 \pm 0.002$ GeV, $b=0.139 \pm 0.016$ GeV$^{-1}$,
$c=0.053 \pm 0.021$ GeV$^{-3}$ and $d$ and $e$ are given as:

\begin{center}
\begin{table}[!htb]
\label{tablesimple}
\begin{tabular}{|c|c|c|}
\hline
    & $d[GeV]$  & $e[GeV^{-1}]$ \\
\hline
$B$ & 1.308(28) & 0.273(8)      \\
\hline
$Q$ & 0.0211    & 0.106         \\
\hline
$S$ & 0.214     & 0.161         \\
\hline
\hline
\end{tabular}
\end{table}
\end{center}

\noindent
where $B$, $Q$ and $S$ indicates the values of $d$ and $e$ for the
corresponding chemical potential given in Eqn.\ref{eq.energy}.

\begin{figure}[!htb]
 \includegraphics[scale=0.25,angle=270]{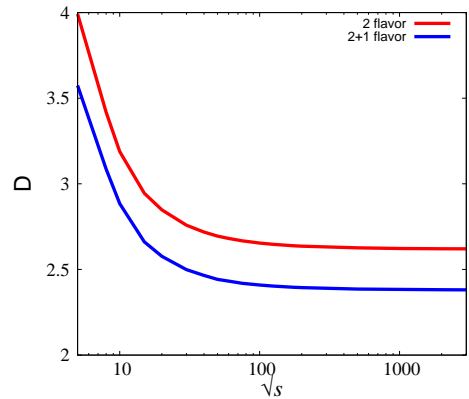}
\caption{(color online) $D$ as a function of $\sqrt{s}$  computed along
          the freeze-out curve.}
\label{fg.ds}
\end{figure}

We have thus calculated $D$ as a function of $\sqrt{s}$ along the
freeze-out curve. Results are shown in Fig. \ref{fg.ds}. We have
varied $\sqrt{s}$ from $5$ GeV to $3$ TeV. $D$ picks up a value of
$\sim 4$ for 2 flavor and $\sim 3.5$ for 2+1 flavor at the low
$\sqrt s$, drops down to a value of 2.6 for 2 flavor and 2.4 for 2+1
flavor around $\sqrt{s} \sim 200$ GeV, and saturates at these values
even for increasing $\sqrt s$. It is highly exciting to find that
the general features of $D$ vs $\sqrt s$ curve obtained in the PNJL
model are found to be similar to those obtained directly in heavy-ion
collision experiments as shown in Fig.4 of Ref.\cite{alice}.
Furthermore the numerical range of $D$ itself is exceptionally
consistent.

It should however be remembered that $D$ as given in Fig. \ref{fg.ds}
is the value obtained when the system is in complete thermal
equilibrium at the given values of temperature and chemical potentials.
Since here we are on the freeze-out curve, we are always inside the
hadronic phase, i.e. the whole of the curve in Fig. \ref{fg.ds} is
corresponding to varying environmental conditions in the hadronic
phase. Thus if our results become completely consistent with 
experimental results the outcome will be that there is no signature of
partonic phase in $D$. At present it seems that the results for STAR
data are above and those of ALICE data are below our model curve. A
more concrete analysis would be possible once the complete experimental
data for $D$ are published.

To summarize, we presented here model study of net charge fluctuations
in terms of $D$-measure using the PNJL model. We found that $D$ does
not have a clear order parameter like behavior. However, given the
temperature and the chemical potentials it is possible to estimate
the corresponding $D$ in PNJL model and compare with experiments. 
Such a preliminary comparison has been done in this work giving
encouraging results and indicating the possibility of detecting
signatures of exotic phases in heavy-ion collisions.

This work is funded by CSIR, UGC(DRS $\&$ UPE) and DST. We thank Tapan
Nayak, Anirban Lahiri and Sarbani Majumder for useful discussions.

\end{document}